\documentclass[prl,twocolumn,amsmath,showpacs]{revtex4}

\usepackage{graphicx}
\usepackage{dcolumn}

\begin{document}

\title{Elucidation of the electronic structure of semiconducting single-walled
 carbon nanotubes by electroabsorption spectroscopy}

\author{Hongbo Zhao}
\altaffiliation[Current address: ]
 {Department of Physics, University of Hong Kong, Hong Kong, China}

\author{Sumit Mazumdar}
\affiliation{
  Department of Physics, University of Arizona, Tucson, Arizona 85721, USA}
\date{\today}

%
%
%

%
%
\begin{abstract}
We report benchmark calculations of electroabsorption in
semiconducting single-walled carbon nanotubes to provide motivation
to experimentalists to perform electroabsorption measurement on these
systems. We show that electroabsorption can detect continuum bands
in different energy manifolds, even as other nonlinear absorption measurements
have failed to detect them. Direct determination of the binding
energies of excitons in higher manifolds thereby becomes possible.
We also find that electroabsorption can provide evidence
for Fano-type coupling between the second exciton and the lowest
continuum band states.
\end{abstract}

\maketitle

Semiconducting single-walled carbon nanotubes (S-SWCNTs) are being
intensively investigated because of their unique properties and broad
potential for applications \cite{McEuen02,Chen05}.  Recent theoretical
investigations have emphasized the strong role of
electron-electron interactions and the consequent excitonic
energy spectra in S-SWCNTs
\cite{Ando,Spataru,Chang,Perebeinos,Zhao}.
While within one-electron theory
two-photon absorption (TPA) begins at the same energy threshold as
the lowest one-photon
absorption, exciton theories
predict a significant energy gap between the lowest two-photon
exciton and the optical exciton \cite{Maultzsch05,Zhao06}.
This energy gap gives the lower
bound to the binding energy of the lowest exciton, and
has been determined experimentally
\cite{FWang05,Maultzsch05,Dukovic05,Zhao06}.
Exciton theories of S-SWCNTs therefore may be considered to have firm footing.

We note, however, that existing experiments have focused almost entirely on the
lowest exciton and
information on the {\it overall energy spectra}
of S-SWCNTs is severely limited. To begin with, no signature of
even the lowest continuum band is obtained from such experiments.
Equally importantly,
finite diameters of S-SWCNTs lead to subband quantization and consequently
a series of energy manifolds labeled $n$ = 1, 2, ..., etc.\ with increasing
energies (see Fig.~4 in Ref.~\cite{Zhao06}),
and although emission studies detect the optical
exciton in the $n=2$ manifold ($Ex2$) \cite{Bachilo02},
it has not been possible to determine its binding energy.
TPA or transient
absorption techniques used to determine the binding energy of the optical
exciton in the $n=1$ manifold ($Ex1$)
are not useful for this purpose, as nonlinear absorptions to states in the
$n=2$ manifold will be masked by the strong linear
absorption to $Ex1$ (this is particularly true here as the
energy of $Ex1$ is nearly half that of the states in the $n=2$ manifold).
Interference effects
between $Ex2$ and the $n=1$ continuum band, suggested from
relaxation studies of $Ex2$ \cite{Manzoni05,Zhu06},
are also difficult to verify directly.
Clearly, measurements
that can probe much broader energy regions of S-SWCNTs are called for.

In the present Letter, we propose electroabsorption (EA), which
measures the difference between the absorption $\alpha(\omega)$
with and without an external static electric field,
as the ideal technique for
understanding the overall energy spectra of S-SWCNTs.
EA has provided valuable information on both
conventional semiconductors \cite{Aspnes}
and $\pi$-conjugated polymers \cite{Weiser,Guo93,Liess97}.
The similarity in the energy spectra
of $\pi$-conjugated polymers and S-SWCNTs \cite{Zhao,Zhao06}
makes EA particularly attractive. EA spectroscopy of S-SWCNTs has already
been attempted \cite{Kennedy05}, while continuous wave photomodulation
spectroscopy has been interpreted as electroabsorption caused by
local electric fields \cite{Gadermaier06}.
EA measurements are currently
difficult as complete separation of semiconducting from metallic
SWCNTs has not been possible to date.
Recent advances in the syntheses of chirality enriched S-SWCNTs \cite{separation}
strongly suggest that EA measurements will become possible in select
S-SWCNTs in the near future.
We present here benchmark calculations of EA for several wide nanotubes
that give new insights to their electronic structures,
and provide the motivation for
and guidance to experimental work.

As in our previous work \cite{Zhao,Zhao06} we choose the
semiempirical $\pi$-electron Pariser-Parr-Pople (PPP) model \cite{PPP}
as our field-free Hamiltonian,
\begin{equation}
\begin{split}
\label{ppp}
H_0 = & -\sum_{\langle ij \rangle,\sigma}t_{ij}
         (c_{i\sigma}^\dag c_{j\sigma} + c_{j\sigma}^\dag c_{i\sigma})
      + U \sum_i n_{i\uparrow}n_{i\downarrow} \\
    & + \frac{1}{2} \sum_{i\neq j} V_{ij}(n_i-1)(n_j-1) .
\end{split}
\end{equation}
Here $c^\dagger_{i \sigma}$ creates a $\pi$-electron with spin
$\sigma$ 
on the $i$th carbon atom, $\langle ij\rangle$ implies nearest
neighbors, $n_{i \sigma} = c^\dagger_{i \sigma}c_{i \sigma}$ is
the number of $\pi$-electrons with spin $\sigma$ on the atom $i$, and
$n_i = \sum_{\sigma} n_{i\sigma}$ is the total number of $\pi$-electrons on the
atom.
The parameter $t_{ij}$ is the one-electron 
hopping
integral, $U$ is the repulsion between two $\pi$-electrons
occupying the same carbon atom, and $V_{ij}$ the intersite Coulomb interaction.
We have chosen Coulomb and hopping parameters as in
our recent work \cite{Zhao}.
In principle, we should also include
the electron-phonon interactions \cite{Perebeinos}, since EA spectra
for real materials will contain signatures arising from phonon
sidebands \cite{Weiser}.
This, however, will make the EA calculations much too complicated.
In agreement with prior EA experiments on $\pi$-conjugated polymers
\cite{Weiser}, our calculations here find that EA signals due to excitons
and continuum bands are sufficiently different that
there can be no confusion in distinguishing between exciton sidebands
and continuum bands.

We are interested in optical absorptions polarized parallel to the
nanotube axis and consider only the component of the static electric field
along the same direction.
The overall Hamiltonian
is written as \cite{Guo93}
\begin{equation}
  H = H_0 + e F z = H_0 + \mu F,
  \label{eq:H}
\end{equation}
where $e$ is the charge of the electron, $F$ the field strength along
the nanotube axis (taken to be the z-direction), and $\mu$ the
transition dipole operator along z.

The EA is calculated in two steps \cite{Guo93}. We first diagonalize
$H_0$ in the space of all single excitations from the Hartree-Fock
ground state, using the single-configuration interaction (SCI)
approximation \cite{Zhao}.
Eigenstates of S-SWCNTs
are of even ($A_g$) or odd ($B_u$) parity, and dipole
matrix elements are nonzero only between states of opposite
parity \cite{note}.
We calculate the
field-free absorption spectra $\alpha(\omega;0)$ from the calculated
dipole matrix elements between the ground 1$A_g$ state and excited $B_u$ states
\cite{Zhao,Zhao06}.  We now evaluate the matrix elements of $\mu$
between {\it all} excited states of $H_0$, construct and diagonalize the
total Hamiltonian $H$ with the eigenstates of $H_0$ as the basis
states, and calculate the new absorption $\alpha(\omega;F)$.
The EA is given by
   $\Delta \alpha(\omega;F) = \alpha(\omega;F)-\alpha(\omega;0)$.

The effect of the nonzero field is to mix $A_g$ and $B_u$ states.
Within second order perturbation theory appropriate for weak fields,
$Ex1$ undergoes a Stark energy shift $\Delta E_{Ex1}$, given by
\begin{equation}
  \Delta E_{Ex1} = \sum_j\frac{|\langle Ex1 | \mu | jA_g \rangle|^2\,F^2}
                             {E_{Ex1}-E_{jA_g}} ,
  \label{eq:dE}
\end{equation}
where the sum over $A_g$ states includes the $j=1$ ground state.
In addition, $A_g$ excitons that are forbidden for $F=0$ become weakly
allowed for $F\neq0$. This transfer of oscillator strengths between
$A_g$ and $B_u$ excitons is also quadratic in $F$. Nondegenerate perturbation theory
cannot, however, describe the mixing of $A_g$ and $B_u$ states belonging to
continuum bands and the EA in these energy regions can be only
calculated numerically. As we discuss below, the same is true for
$Ex2$, which is buried within the $n=1$ continuum. We have ignored
the dark excitons \cite{Perebeinos,Zhao} in our discussion, as they
play no role in linear or nonlinear absorption.



\begin{figure}[tb]
  \includegraphics[width=3.37in]{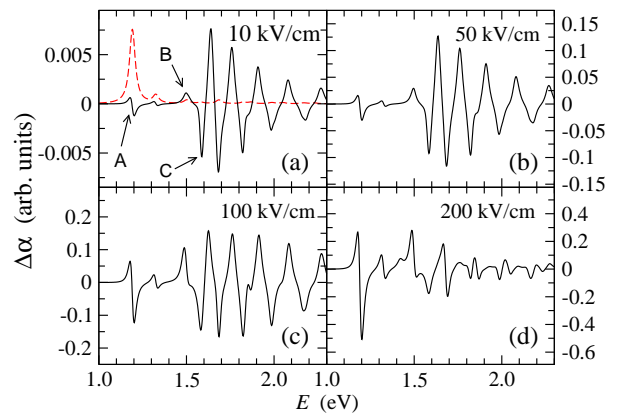}
  \caption{\label{f:EA1}
    (Color online) (a) Linear absorption (red), and EA
    spectrum of the (10,0) S-SWCNT in the $n=1$ energy region for $F$ = 10 kV/cm.
    (b)--(d) EA spectra for $F$ = 50, 100, and 200 kV/cm, respectively.
}
\end{figure}
We first describe the $n=1$ energy region separately.
In Fig.~\ref{f:EA1}(a) we have plotted the calculated
linear absorption
along with EA spectrum in the energy range corresponding to
the $n=1$ manifold for $F=10$ kV/cm for the (10,0) S-SWCNT.
Figs.~1(b)--(d) show the EA spectra for $F=50$, 100, and 200 kV/cm, respectively.
EA for other S-SWCNTs, including chiral ones, are similar.
The three most
important features of the EA spectra are indicated in
Fig.~\ref{f:EA1}(a). The derivative like
feature~A corresponds to the redshift of $Ex1$.
From Eq.~(3), the redshift (as opposed to a blueshift) is the consequence
of the existence of an $A_g$ two-photon exciton that is closer in energy to $Ex1$
than the 1$A_g$ ground state,
and that also has a stronger dipole coupling to $Ex1$ \cite{Guo93}.
In analogy to $\pi$-conjugated polymers \cite{Guo93}, we have previously
referred to the two-photon exciton state as the $mA_g$ \cite{Zhao06}.
It is this state that is visible in TPA and transient absorption
\cite{FWang05,Maultzsch05,Dukovic05,Zhao06}.
Feature~B in Fig.~1(a) corresponds to the field-induced absorption to the
$mA_g$.
Feature~C is a dip in the absorption due to the $B_u$ state
at the threshold
of the
$n$ = 1 continuum band (hereafter the $nB_u$ \cite{Zhao06}). The continuum band
is recognized by its oscillatory nature, and its appearance over a broad energy
region where there is no linear absorption.
EA can therefore give the
binding energy of $Ex1$ directly, as the energy difference between the
features C and A.

The amplitude of the continuum band signal in Fig.~1 is much larger
than that of the exciton at low field. This has been observed
previously in a crystalline polydiacetylene
\cite{Weiser}.
Finite conjugation lengths prevent the
observation of the continuum band EA signal in disordered
noncrystalline $\pi$-conjugated polymers \cite{Liess97}, but this
signal will be observable in S-SWCNTs where the tubes are known to be long.
Features due to the exciton and the continuum can be distinguished
easily even when electron-phonon interactions lead to sidebands,
from their different field
dependence \cite{Weiser}.
In Fig.~2(a) we show that the calculated field dependence of both the
energy shift of $Ex1$
and the amplitude of EA signal due
to $Ex1$ for (6,5), (7,6), and (10,0) S-SWCNTs are quadratic in $F$
up to the largest value of $F$. EA amplitudes due to the $nB_u$,
also plotted in the figure, exhibit weaker dependence on $F$ at strong
fields, in agreement with that observed in the crystalline
polydiacetylenes \cite{Weiser}.
In Figs.~2(b) and (c) we have plotted EA signals due to $Ex1$ and the $nB_u$,
respectively, at different fields.
EA due to the
continuum (but not the exciton) exhibits the expected band-broadening
\cite{Weiser} as a function of the field.


\begin{figure}[tb]
  \includegraphics[width=3.375in]{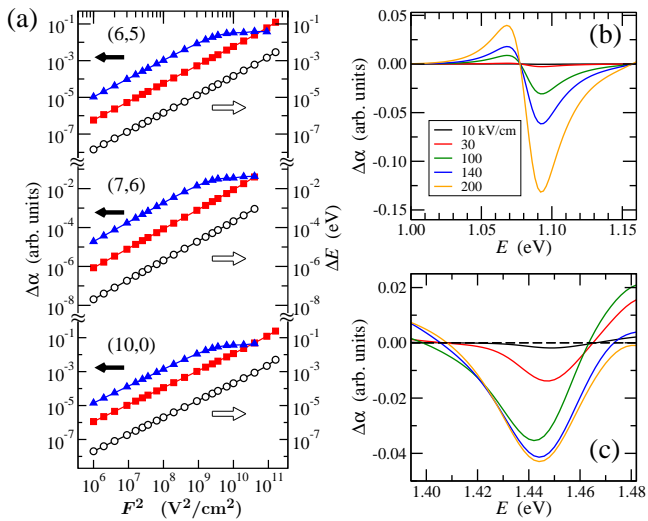}
  \caption{\label{f:Fdep}
    (Color online)
    (a) Field dependence of $\Delta E_{Ex1}$ (open circles), the EA amplitude of
     $Ex1$ (solid squares) and $nB_u$ (solid triangles) for the (6,5), (7,6),
     and (10,0) S-SWCNTs.
    (b) and (c) Field dependence of EA signals for the (6,5) S-SWCNT
     due to the $Ex1$ and the $nB_u$, respectively.
  }
\end{figure}

We now discuss EA over the entire energy region, focusing on the
$n=2$ manifold.  In Fig.~3(a) we have shown the calculated linear absorption
for the (6,5) S-SWCNT, while Fig.~3(b) shows the EA spectrum for $F=50$ kV/cm.
The EA spectrum is dominated by two distinct and slowly
decaying oscillating signals due to the $n=1$ and 2 continuum bands.
Comparison of Figs.~3(a) and (b) indicates that $Ex2$ lies within the $n=1$ continuum.
We also note that the signal due to $Ex2$
is much stronger than that due to $Ex1$.
This is a consequence of the interference between $Ex2$ and the $n=1$ continuum,
as we prove by comparing the true EA of Fig.~3(b) with that in Fig.~3(c), where
the coupling between $Ex2$ and the $n=1$ continuum states
has been artificially eliminated.
We calculated EA from the $n=1$ manifold states alone by removing all
$n=2$ states from Eq.~(2); similarly the calculation of EA from the $n=2$ states
ignored the $n=1$ states. Fig.~3(c) shows the superposition of these two independent EAs.
The very small EA signal due to $Ex2$ in Fig.~3(c) is completely obliterated
by the smooth envelope of the $n=1$ continuum, clearly indicating that the much
larger signal in Fig.~3(b) is a consequence of the coupling between $Ex2$ and
the $n=1$ continuum states. We have further verified this by performing the
EA calculations with varying $U/t$: larger Coulomb interactions imply larger
Coulomb coupling between $Ex2$ and the $n=1$ continuum, and give larger EA signal
for $Ex2$.
We have observed these characteristics
in our calculated EA spectra for all S-SWCNTs.


\begin{figure}[t]
  \includegraphics[width=3in]{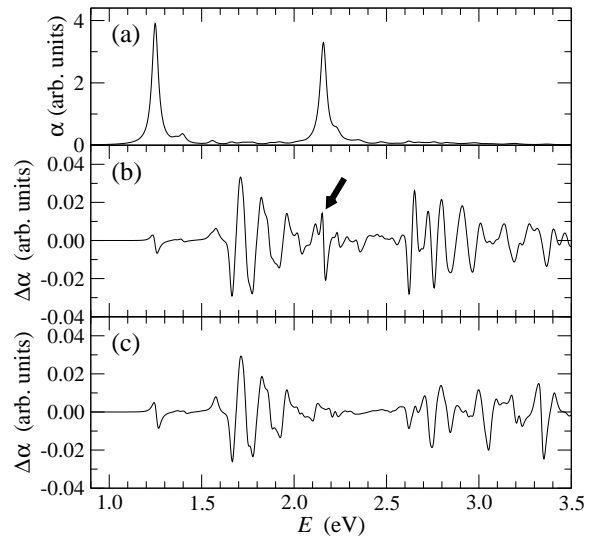}
  \caption{\label{f:EA2}
    Linear absorption (a) and
    EA spectrum (b) in the energy region covering both $n=1$ and 2 energy manifolds for
    the (6,5) S-SWCNT for $F=50$ kV/cm. The arrow indicates $Ex2$.
    (c) Superposition of the EAs for
    the $n=1$ and 2 energy manifolds, calculated separately and independently of one another, at the same field.
  }
\end{figure}

Coulomb coupling between a discrete state with a continuum
leads to the well-known
Fano resonance, which manifests itself as a sharp asymmetric line in
the {\it linear} absorption \cite{Fano61}. Calculation of the absorption
spectrum of the (8,0) S-SWCNT has previously found this coupling \cite{Spataru}.
In contrast to this standard description of the Fano effect,
the interference effect we observe in the EA spectrum
is a consequence of transition dipole coupling between $Ex2$ and the
$n=1$ continuum states.
One interesting consequence of this dipole coupling is that unlike
$Ex1$, which undergoes redshift in all cases, $Ex2$ can be either
redshifted [as observed in our calculations for (6,4), (7,6), and
(11,0) S-SWCNTs] or blueshifted [observed for (6,2), (8,0), and (10,0)
S-SWCNTs].
The reason for this is explained in Table~I, where we have
listed for the (8,0) and the (6,4) S-SWCNTs the dominant transition dipole
couplings between $Ex2$ and $A_g$ states
along with the energy differences between them.
Among these states, only one is from the $n=2$ manifold, which is the $mA_g2$ state,
the equivalent of the $mA_g$ state in the $n=2$ manifold. All other states belong to the $n=1$ continuum.
The relatively large energy difference between the $mA_g2$ and $Ex2$,
comparable to that between the $mA_g$ and $Ex1$, indicates that the
energy shift of $Ex2$ is {\it
determined predominantly by the dipole-coupled $n=1$ continuum
states}. States below and above $Ex2$ contribute to blue and
red shifts, respectively,
and the energy differences in Table~I
rationalize blue (red) shift in the (8,0) ((6,4)) S-SWCNT.
The {\it magnitude} of the energy shift of $Ex2$ in all cases is
smaller than that of $Ex1$ because of partial cancelations,
even as the amplitude of the EA signal of $Ex2$ is larger.

\begin{table}[tb]
  \caption{\label{t:couple}
    Dominant transition
    dipole couplings between $Ex2$ and
    $A_g$ states,
    as well as the corresponding energy differences.
    The $mA_g2$ state is labeled with an asterisk (*).
  }
  \begin{ruledtabular}
  \begin{tabular}{cD{.}{.}{12}D{.}{.}{8}}
    ($n$,$m$)
    & \multicolumn{1}{c}{$\langle Ex2|\mu| jA_g\rangle / \langle Ex2|\mu| 1A_g\rangle$}
    & \multicolumn{1}{c}{$E_{Ex2}-E_{jA_g}$ (eV)}      \\ \hline
    (8,0) & 8.31^* & -0.423 \\
          & 8.01   &  0.057 \\
          & 6.22   & -0.103 \\
          & 1.22   &  0.073 \\[0.6mm]
    (6,4) & 15.3   &  0.049 \\
          & 11.5   & -0.051 \\
          & 11.1   &  0.081 \\
          & 9.64   & -0.070 \\
          & 9.45   & -0.080 \\
          & 6.62   & -0.025 \\
          & 6.45^* & -0.402
  \end{tabular}
  \end{ruledtabular}
\end{table}

\begin{table}[b]
  \caption{\label{t:comp}
    Relative weights of Hartree-Fock $n=1$ excitations in the
    SCI $Ex2$ eigenstate of several S-SWCNTs.
  }
  \begin{ruledtabular}
  \begin{tabular}{cccccccc}
               & (8,0) & (10,0) & (6,2) & (6,4) & (6,5) & (7,6) & (9,2) \\ \hline
   percentage  &  3\%  &  2\%   &  23\% &  12\% &  20\% &  33\%  & 26\%
  \end{tabular}
  \end{ruledtabular}
\end{table}

Correlated SCI eigenstates of the Hamiltonian (1) are superpositions of
band-to-band excitations from the Hartree-Fock ground state.
Furthermore, within the noninteracting tight-binding model as well as within
Hartree-Fock theory, matrix elements of the component
of the transition dipole moment along the nanotube axis are nonzero only
for ``symmetric'' excitations, viz.,
from the highest valence band to the lowest conduction band,
from the second highest
valence band to the second lowest conduction band, etc.
Hence the strong
dipole couplings between $Ex2$ and proximate $n=1$ continuum eigenstates of
$A_g$ symmetry {\it necessarily implies that $Ex2$ eigenstate contains
basis vector components belonging to both $n=1$ and $n=2$ manifolds}.
This is precisely the signature of Fano coupling.
Table~II shows the relative weights of the $n=1$
one electron-one hole excitations in the correlated $Ex2$ eigenstates of
several S-SWCNTs. These contributions are chirality-dependent,
and reach as high as 30\%.

The EA due to the $n=2$ continuum in Fig.~3(b) is similar
to that of the $n=1$ continuum.
The threshold of the $n=2$ continuum is always detectable
in our calculated EA spectra. Further confirmation of the band
edge can come from measurements of its field-dependence, which is the
same as for $n=1$. Taken together with emission measurements that give
the energy location of $Ex2$ \cite{Bachilo02},
EA can then give the precise binding energy of $Ex2$.

In conclusion, we have performed benchmark calculations
showing
that EA measurements in S-SWCNTs can provide valuable information
on their electronic structures that are
difficult to obtain from other measurements. In particular,
EA spectroscopy can detect both $n=1$ and 2 continuum bands.
EA due to continuum bands can be easily differentiated from those due to
excitons.
Precise estimates of the binding energies of both $Ex1$ and
$Ex2$ can therefore be obtained. $Ex1$ is predicted to have a redshift, which would
provide indirect evidence for the $mA_g$ state detected in
complementary nonlinear absorption measurements
\cite{FWang05,Dukovic05,Maultzsch05,Zhao06}. High resolution
will also allow direct detection of the $mA_g$.
We find strong evidence for
Fano-type coupling between $Ex2$ and $n=1$ continuum states that has previously
been suggested from the experimental observation
of ultrafast nonradiative relaxation of $Ex2$ \cite{Manzoni05,Zhu06}.
EA measurements can provide a more direct evidence for this.
Syntheses of chirality enriched S-SWCNTs are allowing
a variety of sophisticated spectroscopic measurements. The results reported
here provide strong motivation for EA spectroscopy of S-SWCNTs.
It is tempting to extend the
current EA calculations to the $n=3$ energy region. Although the $n=3$
exciton is identifiable from linear absorption calculations, our calculations
indicate that its
coupling to the $n=2$ continuum states is even stronger than that of $Ex2$
with the $n=1$ continuum. The analysis of
excitons and continuum bands become rather complicated at these high energies. Work is
currently in progress along this direction.

This work was supported by NSF-DMR-0406604.


\begin{thebibliography}{99}
\bibitem{note}
  This is only approximately true for chiral NTs,
  but in practice there is no difference between zigzag and chiral;
  see Ref.~\cite{Zhao}.
\end{thebibliography}
\end{document}